# Optical Synthesis of Transient Chirality in Achiral Plasmonic Metasurfaces

*Andrew S. Kim[1], Anjan Goswami[1], Mohammad Taghinejad[1,2] and Wenshan Cai[1,3]*\*

[1] School of Electrical and Computer Engineering, Georgia Institute of Technology, Atlanta, Georgia 30332, USA

[2] School of Materials Science and Engineering, Stanford University, Stanford, California 94303, USA

[3] School of Materials Science and Engineering, Georgia Institute of Technology, Atlanta, Georgia 30332, USA

\*E-mail: wcai@gatech.edu

**Abstract**

As much as chiral metasurfaces are significant in stereochemistry and polarization control, tunable chiroptical response is important for their dynamic counterparts. A single metasurface device with invertible chiral states can selectively harness or manipulate both handedness of circularly polarized light upon demand, where in fact chiral inversion in molecules is an active research field. Tactics for chirality switching can be classified into geometry modification and refractive index tuning. However, these generally confront slow modulation speed or restrained refractive index tuning effects in the visible regime with forbidden 'true' inversion. Here, we reconfigure the 'optical' geometry through inhomogeneous spatiotemporal distribution of hot carriers as a breakthrough, transforming a plasmonic achiral metasurface into an ultrafast transient chiral medium with near-perfectly-invertible handedness in the visible. The photoinduced chirality relaxes through the fast spatial diffusion process of electron temperature compared to electron-phonon relaxation, empowering hot-carrier-based devices to be particularly suitable for ultrafast chiroptics.

**Introduction**

The time-dependent rotation of electric fields within propagating light, a key feature of circularly polarized light (CPL), offers several important traits ready to be leveraged. The spin-angular momentum carried by CPL endows the capability to serve as an essential tool in opto-spintronics and modern characterization techniques for solid-state physics. Moreover, the orthogonality between the two CPL states, namely the left-circularly polarized light (LCP) and right-circularly polarized light (RCP), are utilized for communication, displays, data storage, encryption, and quantum computing. Focusing on the geometrical aspect, the helical paths which the electric fields of LCP and RCP follow are non-superimposable mirror images - a property well-known as "Chirality". The chiral nature of CPL gives rise to the capability of resolving the spatial arrangement of atoms and thus discerning the handedness of a chiral molecule which the difference in the interaction between the CPL states and chiral objects manifests itself as circular dichroism (CD)[1,2]. Indeed, discriminating chiral biomolecules is very important as different enantiomers – molecules with same molecular formulas but with different handedness – in medicines are influential to life, which could either be clinically beneficial or cause catastrophic health conditions[3]. That said, easy access to chiroptical effects is of extreme importance due to the capability of both CPL control and chiral detection.

However, such a microscopic length scale of molecules when compared to photon wavelengths makes chiroptical effects inherently weak. In order to obtain strong and easily accessible chiroptical effects, periodically arranged optical nano-resonators with designed geometries have been proposed as the artificial counterpart[4,5,6,7,8], where numerous studies have reported artificial chiroptical effects for different application purposes[9,10,11,12,13,14]. Such class of engineered materials in a two-dimensional configuration is known as chiral metasurfaces. A common downside of metasurfaces is the lack of post-fabrication tunability, in which their optical functionalities are fixed immediately after fabrication based on their geometries and the refractive indices of the constituent materials. Chiral metasurfaces are no exception to this shortcoming. This consequently requires one to prepare complete sets of enantiomorphs in order to fully exploit or manipulate both states of CPL, doubling the fabrication time and cost. Thus, the pursuit of tuning the chiroptical effect, and ultimately inverting the handedness of a chiral metasurface, has long been a companion throughout the history of chiral metasurface design[15]. Novel strategies for tuning the geometrical[16,17,18], optical[19,20,21], and experimental configuration[22] were introduced.

However, each tactic has their own constraints such as intrinsically slow tuning speed or sophisticated fabrication requirements for the case of effective geometry tuning. For refractive index tuning, demonstrations are mostly done outside of the optical regime where 'true' inversion is fundamentally limited due to the geometrical handedness presets of chiral metasurfaces.

Meanwhile, recent studies on photoinduced symmetry breaking[23, 24] hints the possibility to alleviate the aforementioned constraints. Nano-resonators in metasurfaces made of noble metals support localized surface plasmons upon illumination which, shortly after excitation, can non-radiatively decay by generating hot carriers via Landau damping[25, 26, 27]. The perturbed electron energy distribution results in modified complex refractive indices of the material components and thus changes the optical response of the metasurface[28, 29]. The extremely short timescale of hot-carrier generation and relaxation offers ultrafast tuning of optical properties which is a matter of extensive research[30, 31, 32]. Close inspections on hot-carrier dynamics suggest that hot-carrier generation efficiency is proportional to the field intensity. Subsequently, the population of generated hot carriers follow the field enhancement profile (referred to as "hot spots") of the nano-resonators[33]. Hence, engineered near-field profiles under optical excitation allow inhomogeneous hot-carrier distributions.

In this work we demonstrate ultrafast, complete chirality inversion in the visible regime enabled by inhomogeneous hot-carrier generation within each resonator. Within the resonator, the broken optical mirror symmetry due to the refractive index gradient induced by inhomogeneous hot-carrier generation instigates chirality in achiral structures. Through this configuration, desired handedness can be assigned based on the near-field profile of the resonator responsible for hot-carrier excitation. Thus, simple polarization adjustment of input light for hot-carrier generation is all we need to assign the chiral state of the metasurface. Furthermore, the relaxation (homogenization) of the anisotropic refractive index profile of the nano-resonator is dictated by electron temperature spatial diffusion[34, 35], which is a recent addition to the portfolio of strategies for expediting the tuning speed typically limited by the electron-phonon decay for all-optical tuning purposes[24, 36, 37, 38]. This work introduces a novel approach for versatile, ultrafast chirality control which achieves complete chiral inversion in the visible regime. We anticipate this technique to inspire novel approaches for chiral inversion, and to serve as an effective tool for all-optical signal processing, near-field chirality probing, and understanding spatiotemporal dynamics of hot carriers.

**Results**

*Principles and design*

Figure 1a depicts the general working principle and design requirements. A double-layered metasurface with gold (Au) nanostripes at the lower layer and Au triangular split-ring resonators (Tr-SRR) with a side angle of 60° at the top layer is our sample of interest. An isolated Tr-SRR possesses two mirror symmetry planes, where one is parallel to the substrate (horizontal) and the other is perpendicular to both the substrate and the longitudinal direction of the nanostripes (vertical). Note that although the refractive index difference of the substrate and air breaks the horizontal symmetry in principle, the substrate effect is known to be negligible. Tr-SRR supports excitation modes where the field enhancement at one leg can be larger than the other according to the polarization state of the impinging light. Pump pulses with different polarization states can therefore selectively excite one of the legs of the Tr-SRR which breaks the vertical mirror symmetry due to the anisotropic hot-electron population at the initial stage. However, an individual Tr-SRR with asymmetric excitation cannot be considered truly chiral as the horizontal mirror symmetry still remains valid. Here, the nanostripes placed under the Tr-SRRs with subwavelength spacing form meta-molecules with broken horizontal mirror symmetry, thereby morphing the excited metasurface chiral. The two circularly polarized probe pulses then monitor the time evolution of the induced chirality while continuously scanning the time delay between the pump and probe pulses, where enantiomers with opposite handedness are created upon pump illuminations with mirror-flipped polarizations. Figure 1b illustrates the microscopic picture of hot-electron-induced symmetry breaking. The transient circular dichroism ($\Delta$CD) as a function of the time delay (t) can be defined as:

$$\Delta CD~(t) = \Delta T_{RCP}~(t) - \Delta T_{LCP}~(t) \qquad \text{eq. (1)}$$

where $\Delta T$ stands for the change in the optical transmission of the metasurface triggered by hot-carrier excitation. The time evolution of $\Delta T$ and $\Delta CD$ can be divided into several regimes. Shortly after the arrival of the pump pulse, nonthermal electrons – electrons of which do not obey the Fermi-Dirac energy distribution – with high energies are generated at the hot spots of the Tr-SRR (note that relatively low field enhancement makes hot-carrier effects from the nanostripes negligible). Subsequently, nonthermal electrons quickly thermalize through electron-electron scattering and an elevated electron temperature with a spatial gradient forms[39, 40], inheriting the inhomogeneity from the spatial distributions of nonthermal electrons. Hence, a refractive index gradient forms and creates chirality. While the hot-electron relaxation process remains active, the meta-molecule loses its chirality when the spatial electron temperature distribution homogenizes via spatial diffusion. The geometric dimensions of the resonators make the nonlocal relaxation timescale faster than the local hot-electron relaxation speed, rendering transient chirality as a rapid version of all-optical switching. Once homogenized, we are only left with the temporal $\Delta T$ response which decays dominantly through the electron-phonon scattering process. The excess lattice heat transferred from energetic electrons gets released to the environment via phonon-phonon scattering and the excited nano-resonators revert to its original state. Figure 1c further describes details of the sample fabricated on glass substrate (see Methods), where the two layers of nanostripes and Tr-SRRs are separated by 40 nm. Unlike isolated structures with mirror symmetry, the spatial invariance of nanostripes in the direction perpendicular to the mirror axis removes misalignment-induced intrinsic chirality occurring from fabrication. Both layers of plasmonic nanostructures are apparent in a scanning electron microscopy image of the fabricated structure given in Figure 1d. As graphically described, we choose the mirror axis as the reference for the polarization angle of the incident light. The measured static optical response of the fabricated sample in Figure 1e indeed shows minimal intrinsic chirality, which couldn't be completely removed due to fabrication imperfection. The transmission dip originating from the Tr-SRR (Supplementary Notes 1) is labeled as $\lambda_{CP}$ (702 nm).

*Hot electron induced chirality*

A center wavelength of 880 nm was chosen for the pump wavelength as undesired pump light detection does not overshadow the probe signal with a wavelength range of 420-800 nm. Furthermore, asymmetric hot spot distribution with large absorption at the specified pump wavelength makes it suitable for characterizing the transient dynamics of the structure. We first start the characterization with a mirror-symmetric pump polarization state which produces symmetric field distributions (Figure 2a) as seen in simulation results. The measured transient responses of the sample are expressed in $\Delta OD$ values where $\Delta OD = -\log_{10}(T/T_0)$. $T_0$ is the static transmission and T is the transmission at specified time delays. In the two transient maps plotted in Figure 2b, we see spectral features of the modified interband transition rates near the d-band transition threshold of Au (~550 nm), and red-shifting plasmonic resonances accompanied by spectral broadening near $\lambda_{CP}$. Although the transient response for both RCP and LCP probes should be identical in principle, we see non-zero discrepancies near $\lambda_{CP}$ due to the intrinsic chirality of the sample. This becomes more evident once we plot the two-dimensional ΔCD map as in Figure 2c. The cross-section of the ΔCD map plotted below (at the time delay that gives maximum ΔCD for asymmetric excitations) implies the ΔCD signal being a modulation feature of the intrinsic chiral response, as spectral features of ΔCD and intrinsic CD are located within a spectral vicinity (rightmost graph).

Moving onto asymmetric near-field excitations, we first excite the left leg of the Tr-SRR with a +50° input polarization (Figure 2d). As we see in Figure 2e, a stronger spectral broadening of the plasmonic resonance for LCP is apparent and we see a somewhat different ΔCD profile compared to the symmetric excitation case (Figure 2f). Delving into the contribution of spatial inhomogeneity, we decompose the ΔCD profile into two parts which are the dynamic modulation of the intrinsic CD (ΔtCD) and the transiently induced chirality (ΔiCD) (Supplementary Notes 2). Despite intrinsic nonlinearities of plasmonic resonances, the perturbative nature of the hot-electron-induced optical response minimizes errors from the linear decomposition interpretation. Further moving on, the mirror-flipped pump polarization direction (-50°) results in inverted field distributions (Figure 2g) and differential transmission profiles (Figure 2h). Contrary to the ΔCD profiles of left and right leg excitations, the ΔiCD profiles of left leg ($\Delta iCD_L$) and right leg ($\Delta iCD_R$) excitation being near-perfect inversions (Figure 2i) validates the interpretation of ΔCD as a collective response of ΔiCD and ΔtCD. Comparing between the ΔCD maps of all three pump excitations, we see that symmetric excitation has slower time dynamics which similar traces can be seen in later delay times in the ΔCD maps of asymmetric excitations as well where sign flipping is even observed due to opposing contributions of ΔiCD and ΔtCD. We deduce that the ΔtCD response is governed by local hot-electron excitation and relaxation processes with minimal contribution from the nonlocal effect, making the process slower than the $\Delta$iCD response.

In order to further understand the characteristics of transient chirality, we start off with analyzing the temporal behavior of the $\Delta$CD response displayed in Figure 3a under +50° pump excitation. We first compare the decay time constants for the $\Delta$OD response for RCP and LCP probes, and the $\Delta$CD response. The time constants were obtained by fitting the experimental values to ex-Gaussian functions where we see a near-identical fit between the experimental and fitted values (Supplementary Notes 3). The $\Delta$CD response already has a shorter lifetime than the $\Delta$OD response. Further decomposing $\Delta$CD into $\Delta$iCD and $\Delta$tCD, we see a fast and slow component where the fast component corresponds to the transiently induced chirality. $\Delta$iCD quickly approaches to an infinitesimal value as $\Delta$tCD approaches the $\Delta$CD value, making $\Delta$iCD a fast component with sub-picosecond response time. $\Delta$tCD on the other hand has a lifetime similar to that of the $\Delta$OD responses, indicating that they share similar modulation mechanisms. We then scan the polarization angle of the pump and track the evolution of the $\Delta$iCD profile. Figure 3b shows the pump polarization dependence of the $\Delta$iCD magnitudes at their maximum for different polarization angles. Note that $\Delta$CD at +90° pump polarization is considered equivalent to the $\Delta$tCD values of other excitations, making $\Delta$iCD as being fixed to zero at +90° which is apparent in the 2D map. The 2D map shows a symmetric spectral response with inverted signs about the +90° polarization axis. Although not completely parallel to the side legs of the Tr-SRR, $\Delta$iCD values are maximized near +45° and +135° (-45°) excitation instead of +30° and +150° (-30°). This is a combined effect of maximizing spatial asymmetry and hot-carrier generation efficiency, given that the generation efficiency at the Tr-SRR is maximized when the input polarization angle is +90° (Supplementary Notes 4). We also plot the spectral profiles of $\Delta$iCD$_{max}(\lambda)$ under varying pump fluences. Unlike typical ΔOD responses, where spectral red-shifts upon increasing pump fluence are observed, we see in Figure 3c a trivial spectral shift in the ΔiCD peaks while increasing the pump fluence. This suggests that the spectral locations of the ΔiCD profile is determined primarily by the effective geometrical asymmetry, and the refractive index change influences more on the ΔiCD magnitude. Plotting the peak magnitude of $\Delta iCD_{max}(\lambda)$ as a function of pump fluence helps estimating the maximally achievable induced chirality, where a good agreement with a linear fit reveals the quasi-linear nature of the transient chirality.

*Functionalities of hot electron induced chirality*

While the sample experiences transient chirality, an input of linearly polarized light is converted to an elliptically polarized state with a rotated long axis due to CD. As depicted in Figure 4a, due to the chiroptical effect, the incident light undergoes different amounts of rotations for varying wavelengths, known as optical rotatory dispersion (ORD). CD and ORD are Kramers-Kronig related quantities each rising from the difference in the imaginary and real part of the circular-polarization-dependent refractive indices. We perform transient stokes measurements in order to characterize the ultrafast polarization rotation (see Methods), where the measured 2D maps of ORD are shown in Figure 4b as a function of probe wavelength and time delay. Note that the measured optical rotation can also be decomposed into components each originating from intrinsic chirality modulation (ΔtORD) and induced chirality (ΔiORD) since the Kramers-Kronig relation is a linear operation. Similar to ΔCD measurements, we focus on ΔiORD since chirality induced by the inhomogeneous hot-electron generation is of our main interest. The spectral profile of ΔiCD and ΔiORD at time stamps with maximum amplitudes are plotted together in Figure 4c, where the qualitative spectral profiles (peak-like and bisignate) and locations satisfies a typical Kramers-Kronig pair. Similar to ΔCD measurements, input pumps with mirror-inverted polarizations (-50° and +50°) result in induced ultrafast polarization rotations with similar amplitude and opposite signs. The measurement signifies the capability of a single metasurface to rotate polarization in opposite directions and with continuous tunability enabled by simple pump polarization adjustments.

The deep correlation between near-field profiles and transient chirality brings connection between the far-field chiral response and near-field chiroptics – an area of study that offers many promising applications[14, 41, 42, 43, 44]. Although mirror symmetric nanostructures possess achiral far-field response, chiral near-field response can still be observed in the form of broken mirror symmetry distributions of hot spots[45] and sign-inverted optical chirality enhancement profiles[46, 47]. Optical chirality is a measure of the twist in the near-field where non-zero net optical chirality results in a far-field chiroptical response of a nanostructure. This can be computed for electromagnetic fields with complex electric **E** field and magnetic **B** field as[48, 49]:

$$C = -\frac{\omega \varepsilon_0}{2} \text{Im}(\mathbf{E}^* \cdot \mathbf{B}) \qquad \text{eq. (2)}$$

where $\varepsilon_0$ is the free space electric permittivity, $\omega$ is the angular frequency of incident light, and $\mathbf{E}^*$ is the complex conjugate of the electric **E** field. Figure 5a shows simulated near-field chiroptical responses including field enhancements and optical chirality enhancements ($\hat{C} = C/|C_0|$, where $C_0$ is the optical chirality of the incident free space CPL) of our sample under LCP ($\hat{C} = -1$) and RCP ($\hat{C} = +1$) illumination. Although local enhancements of optical chirality and chiral hot spot distributions are present in achiral nanostructures, the optical chirality from different regions entirely cancel each other out resulting in achiral far-field response. This prohibits near-field chiral response to be probed with far-field optics, therefore requiring different methods[45, 46, 47] to observe near-field chirality. However, with the use of hot electrons, ultrafast spectroscopy can monitor the near-field chiral response as hot spot distributions under LCP and RCP illumination suffices the requirement for hot-electron-induced chirality. The resulting far-field chiroptical response under LCP and RCP pump is given in Figure 5b, where we see similar results as previously investigated.

**Discussion**

A novel strategy for creating chirality with desired handedness in achiral plasmonic metasurfaces was introduced. The strong spatial asymmetry of hot-electron-induced permittivity distribution originating from the near-field distribution breaks the optical mirror symmetry within the achiral plasmonic nanostructures. Simple adjustment of the pump polarization can induce chirality with near-perfectly-invertible handedness, making the technique a powerful tool for dynamic chiroptical response tuning. Transient chirality especially shows superior switching speed compared to conventional hot-electron-based all-optical switch since electron heat diffusion mechanism exhibits faster relaxation speed than the electron-phonon relaxation rate. Overall, the nonlocal nature of transiently induced chirality makes it possible to further enhance the magnitude, boost the switching speed, or adjust the operation wavelength with ease via judicious geometrical design. Ultrafast optical rotation and near-field chirality probing were explored, showcasing the work as a new addition to conventional techniques utilized for high-speed optical rotation and near field probing. We envision our method to be beneficial in both technological and academic aspects including ultrafast polarization control, active chiroptics, and hot-carrier science.

**Methods**

*Device fabrication*

The double-layered plasmonic structures were fabricated on a glass substrate (Corning, Eagle XG glass) via two runs of aligned e-beam lithography, where each run follows a three-step fabrication process: (i) a standard e-beam lithography process (Elionix ELS G-100) to define the nanopatterns, (ii) e-beam evaporation of 2 nm/40 nm Ti/Au metal using Denton Explorer, and (iii) a lift-off process in acetone to resolve the plasmonic structures. The nanostripes and the alignment marks were formed during the first run, which was then followed by spincoating a transparent dielectric IC1-200 (Futurrex Inc.). In order to ensure the desired spacing (40 nm) between the first and second layer, the spin-on-glass was then etched down to a thickness of 80 nm via reactive ion etching (STS ICP RIE). Subsequently, the Tr-SRR structures were fabricated on top of the spin-on-glass through the second run of e-beam lithography.

*Static optical characterization*

A collimated broadband light from a fiber coupled tungsten halogen lamp (B&W Tek BPS 120, spectral range 250-2600 nm) was used to characterize the broadband optical response of the device. A linear polarizer and a quarter waveplate were used to create LCP and RCP. The transmission spectra of the device at normal incidence were acquired by a homemade spectroscopy system (Supplementary Notes 5). The light was focused onto the device with a lens, and then the transmitted light was collected through an objective lens (Mitutoyo, 20X plan Apo NIR infinity-corrected). The collected light was delivered to the spectrograph system (Princeton Instrument Acton SP 2300i and PIXIS 400B camera), where the transmission of the device was normalized to the transmission through air.

*Ultrafast optical characterization*

The ultrafast pump-probe spectroscopy setup is operated through a regenerative amplified Ti:sapphire femtosecond laser system (Spectra-Physics Solstice Spitfire ACE, 2.5kHz repetition rate, 89.3 fs pulse width, 800 nm wavelength, and a pulse energy of 1.6 mJ per pulse), and the data collection was conducted using a Helios spectrometer (Ultrafast Systems Inc.). An 80:20 beam splitter divides the 800 nm fundamental light from the laser system into two beams. The low intensity part of the fundamental light passes through another 80:20 beam splitter, and the power level gets reduced again by 80%. The laser beam then gets focused on a 2 mm thick sapphire window to generate a white light continuum (WLC) as the probe signal. The generated probe signal passes through a short-pass filter in order to remove the remaining 800 nm pulse after WLC generation. Prior to WLC generation, the low intensity beam first passes through a motorized delay stage in order control the delay between the pump and probe signal, which has a minimum resolution of 30 fs. The probe beam was then focused onto the sample using a parabolic mirror, and the transmitted probe light was collected by an optical fiber coupled into a visible spectrometer. The input polarization of the probe pulse was controlled by a broadband achromatic quarter waveplate (QWP), where a linear polarizer was installed prior to the QWP to ensure a uniform polarization state of the probe signal after WLC generation. The high intensity part is used to operate the optical parametric amplifier (Ultrafast Systems Inc. Apollo). The fundamental light is partially converted into two near-IR pulses, signal and idler, and the combination between the three beams and different nonlinear processes produces single wavelength pump pulses with a desired wavelength. The pump pulse is also focused on the sample through a parabolic mirror, where the pump is incident with a 10° offset from the probe light. The input polarization the pump pulse was adjusted by both broadband achromatic half and quarter waveplate, and the input power of the pump pulse was controlled by an attenuator. An instrument response function value of 150 fs typically returns good numerical fittings for the transient kinetics. Chirp corrections and coherent artifact removal of the obtained transient maps were done via post processing.

*Optical rotation measurements*

The ultrafast optical rotation of light was characterized by installing a liquid crystal (LC) phase retarder (Thorlabs) and a linear polarizer between the sample and the detector. The incident probe light was vertically polarized, which corresponds to 0° based on our notation. We measure the intensity of the transmitted light through the linear polarizer under four different conditions: (β = 0°, 15°, 30°, 45°; α = 0°), where β and α are defined as the direction of the fast axis of the LC and the transmission axis of the linear polarizer, respectively. The induced phase difference between electric field components that are parallel and perpendicular to the fast axis of the phase retarder is notated as δ. The δ values for our LC phase retarder is highly dispersive, which the values as a function of wavelength is given in the Supplementary Information (Supplementary Notes 6). The Stokes parameters (I, M, C, S) were retrieved from the following equation[50]:

$$I_T(\alpha, \beta, \delta) = \frac{1}{2}[I + (M \cos 2\beta + C \sin 2\beta) \cos 2(\alpha - \beta) \qquad \text{eq. (3)}$$
$$+ \{(C \cos 2\beta - M \sin 2\beta) \cos \delta + S \sin \delta\} \sin 2(\alpha - \beta)]$$

And the polarization rotation can be computed through the following relation:

$$\theta = \frac{1}{2} \tan^{-1} \frac{C}{M} \qquad \text{eq. (4)}$$

ASSOCIATED CONTENTS

**Supporting Information**

Supplementary Figures and Supplementary Notes.

AUTHOR INFORMATION

**Corresponding Author**

E-mail: wcai@gatech.edu.

**Notes**

The authors declare no competing financial interest.

**Acknowledgements**

This work was supported in part by the National Science Foundation under Grant No. DMR-2004749, and by the Office of Naval Research under Grants No. N00014-17-1-2555 and No. N00014-19-1-2530 (DURIP). This work was performed in part at the Georgia Tech Institute for Electronics and Nanotechnology, a member of the National Nanotechnology Coordinated Infrastructure (NNCI), which is supported by the National Science Foundation (Grant ECCS-2025462). A.S.K. acknowledges the support of the Department of Defense (DOD) through the National Defense Science & Engineering Graduate (NDSEG) Fellowship Program.

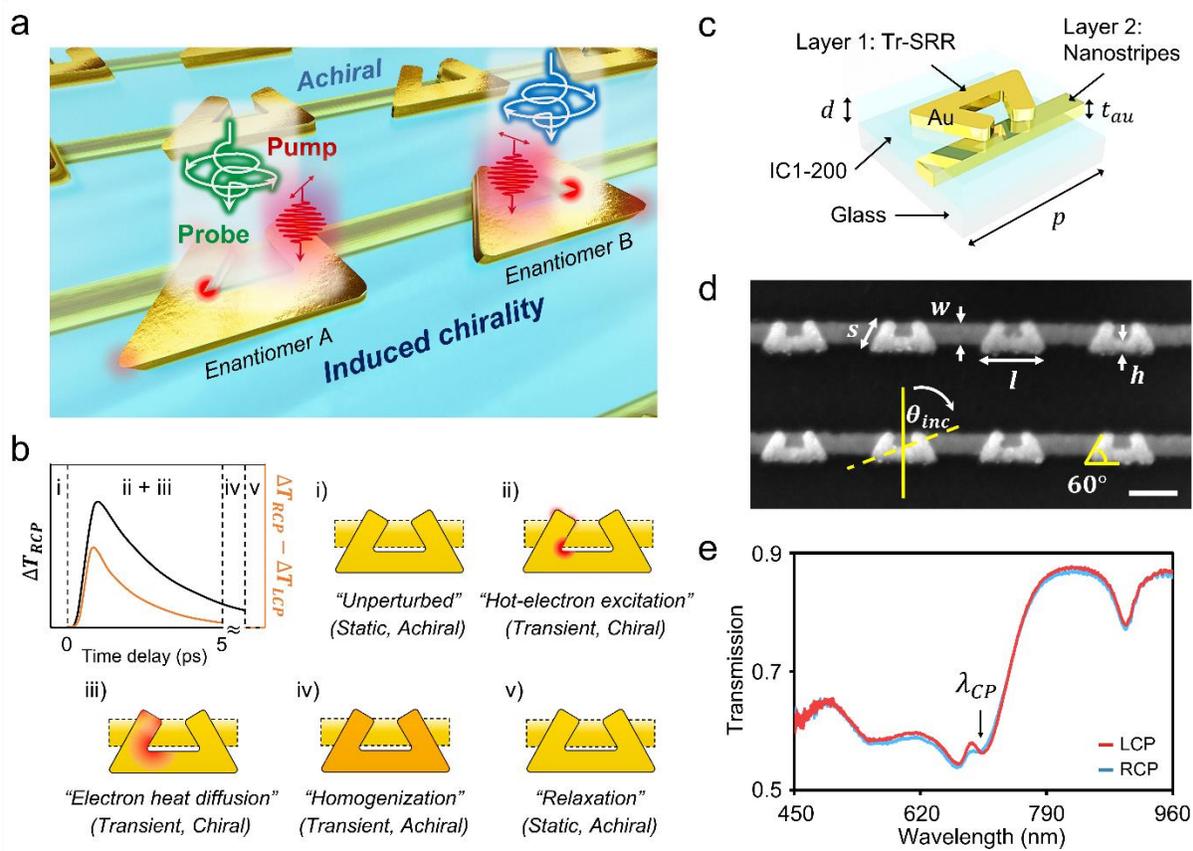

**Figure 1. Concept schematics and sample configurations.** a) Visual sketch of the photoinduced transient chirality. The asymmetric pump-polarization-dependent near-field profile transforms the achiral metasurface into a handedness-selectable chiral object which interacts differently to the left and right circularly polarized light monitored by a probe pulse. b) Conceptual schematic of the transient optical response and the microscopic origin of the transient chirality. The temporal behavior undergoes several phases which the description of the dominating hot-carrier dynamics is illustrated. c) Design of the dual-layered achiral metasurface characterized in this work where the Tr-SRR and the nanostripe is separated by an 80 nm-thick spin-on-dielectric. The dimensions are given as p = 350 nm, d = 80 nm, and $t_{au}$ = 40 nm. d) Scanning electron microscopy image of the sample with further geometrical design specifications: s = 125 nm, w = 85 nm, l = 240 nm, h = 60 nm, and side angle of the Tr-SRR equals to 60°. The center of the Tr-SRR, defined as the center of mass of the triangle that is truncated to form the Tr-SRR geometry, is shifted upwards by 30 nm from the center of the nanostripe. Note that the polarization angle is defined as positive along clockwise rotation since we observe light from behind of the metasurface. The scale bar equals to 200 nm. e) Measured static optical transmission spectra under normal incidence for LCP and RCP light.

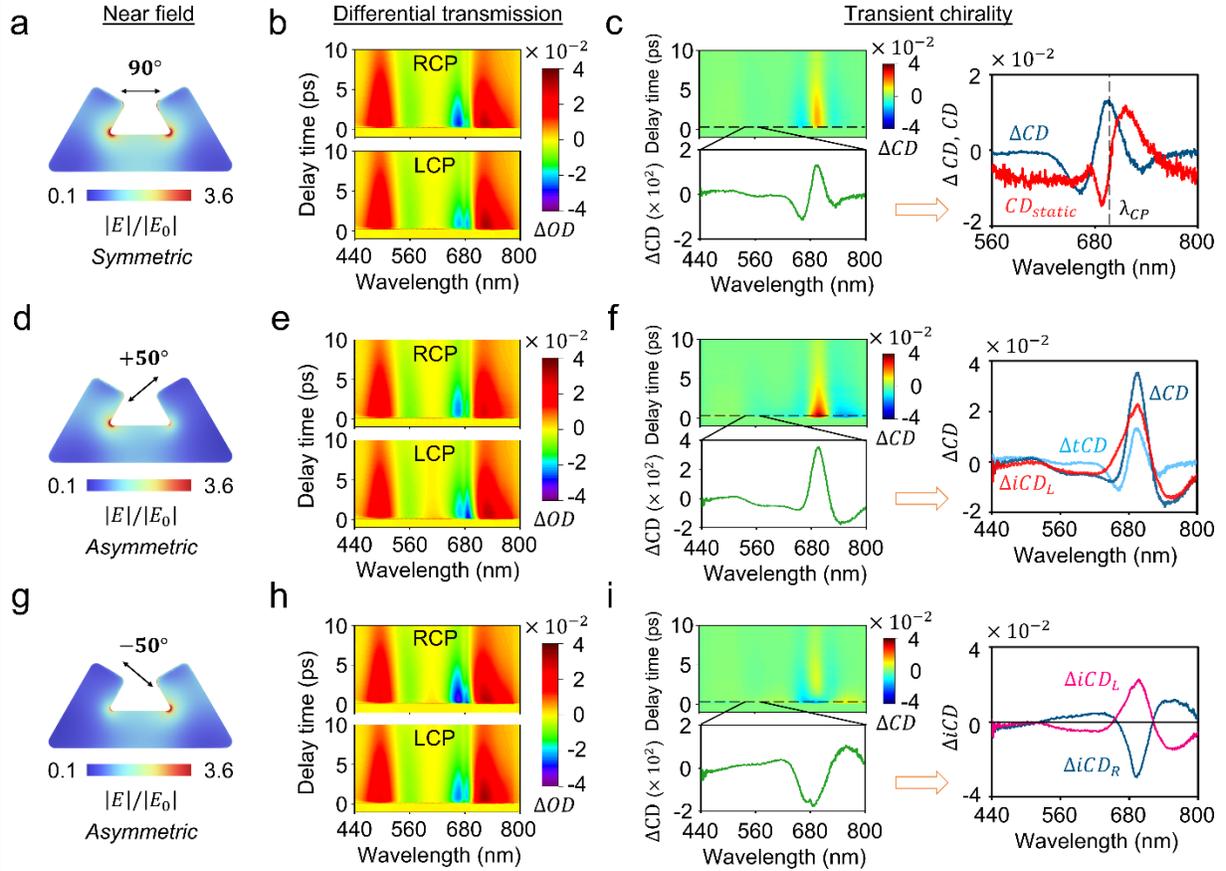

**Figure 2. Transient optical response and induced chirality.** Ultrafast pump-probe spectroscopy measurement results are given under specified pump conditions: energy density of 8 mJ/cm$^2$, duration of 89 fs, and a center wavelength of 880 nm with varying polarization angles. a) Numerically simulated internal electric field enhancement profile of the Tr-SRR under a +90° polarized pump displaying mirror symmetry, where the cross section is taken at the top of Tr-SRR. b) Differential transmission response under RCP and LCP probe (center). c) Measured 2D map of transient circular dichroism as a function of wavelength and delay time (top left), and a 1D cross-sectional view captured at a fixed time delay (bottom left) which is then plotted together with the intrinsic CD profile (right). d-e) Field profile and ΔOD maps under +50° pump polarization selectively exciting the left leg of the Tr-SRR. f) Measured transient chirality response. We decompose ΔCD into components of intrinsic chirality modulation (ΔtCD) and induced chirality

due to left leg excitation ($\Delta iCD_L$). g-h) Results repeatedly acquired under -50° pump polarization selectively exciting the right leg of the Tr-SRR. i) $\Delta CD$ measurements where near-perfect symmetric response between $\Delta iCD_R$ and $\Delta iCD_L$ are observed unlike the overall $\Delta CD$ values which are summated responses of $\Delta iCD$ and $\Delta tCD$.

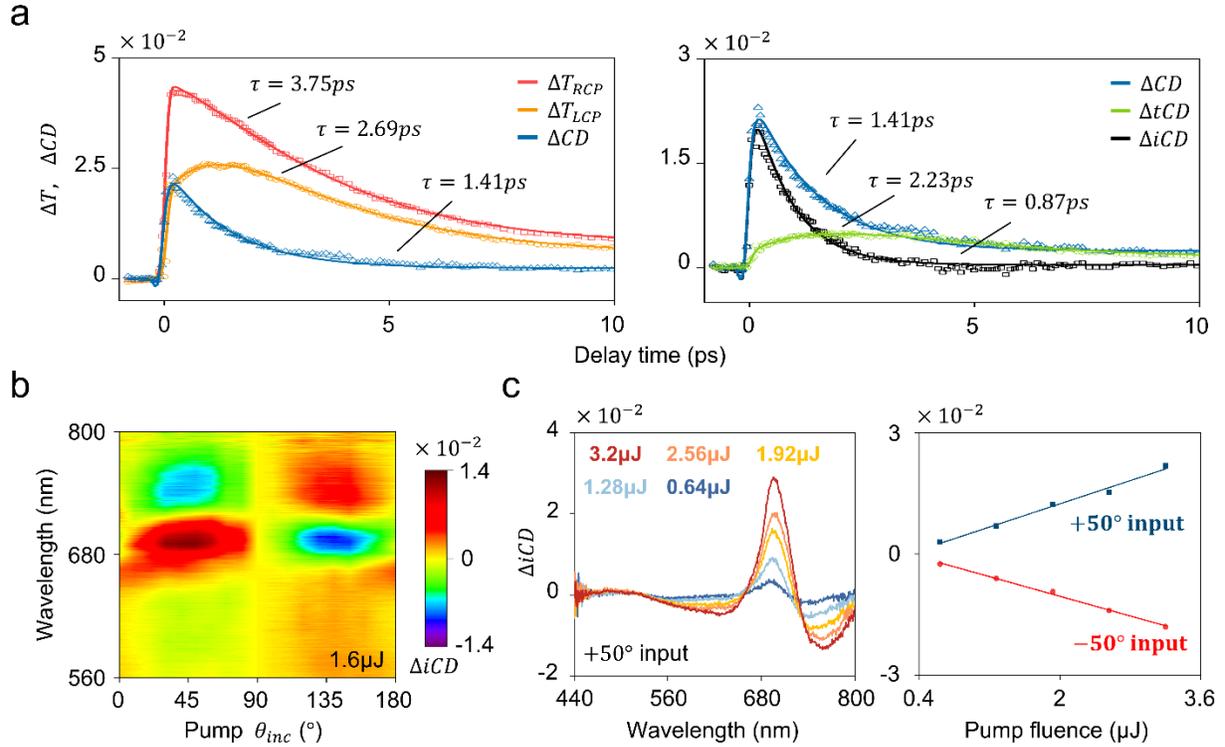

**Figure 3. Characteristics of the induced transient chirality.** a) Transient curves and the recovery time constants of $\Delta T_{LCP}$, $\Delta T_{RCP}$, and the $\Delta CD$ response observed at $\lambda_{CP}$ under +50° pump excitation (left). The $\Delta CD$ curve is then decomposed to temporal profiles of $\Delta iCD$ and $\Delta tCD$, which the decaying timescales suggest the dominating hot-carrier-related mechanism for both features (right). b) Maximum induced chirality spectra as a function of input polarization direction investigated under a pump fluence of 1.6 μJ (corresponding to an energy density of 4 mJ/cm$^2$ in this work). c) Maximum induced chirality spectra for varying pump powers under +50° input (left) and the peak values as a function of pump fluences showing (right). Beyond the pump fluence of 3.2 μJ is approximately the onset of sample degradation due to excessive heating at the hot spots of the nanostructures.

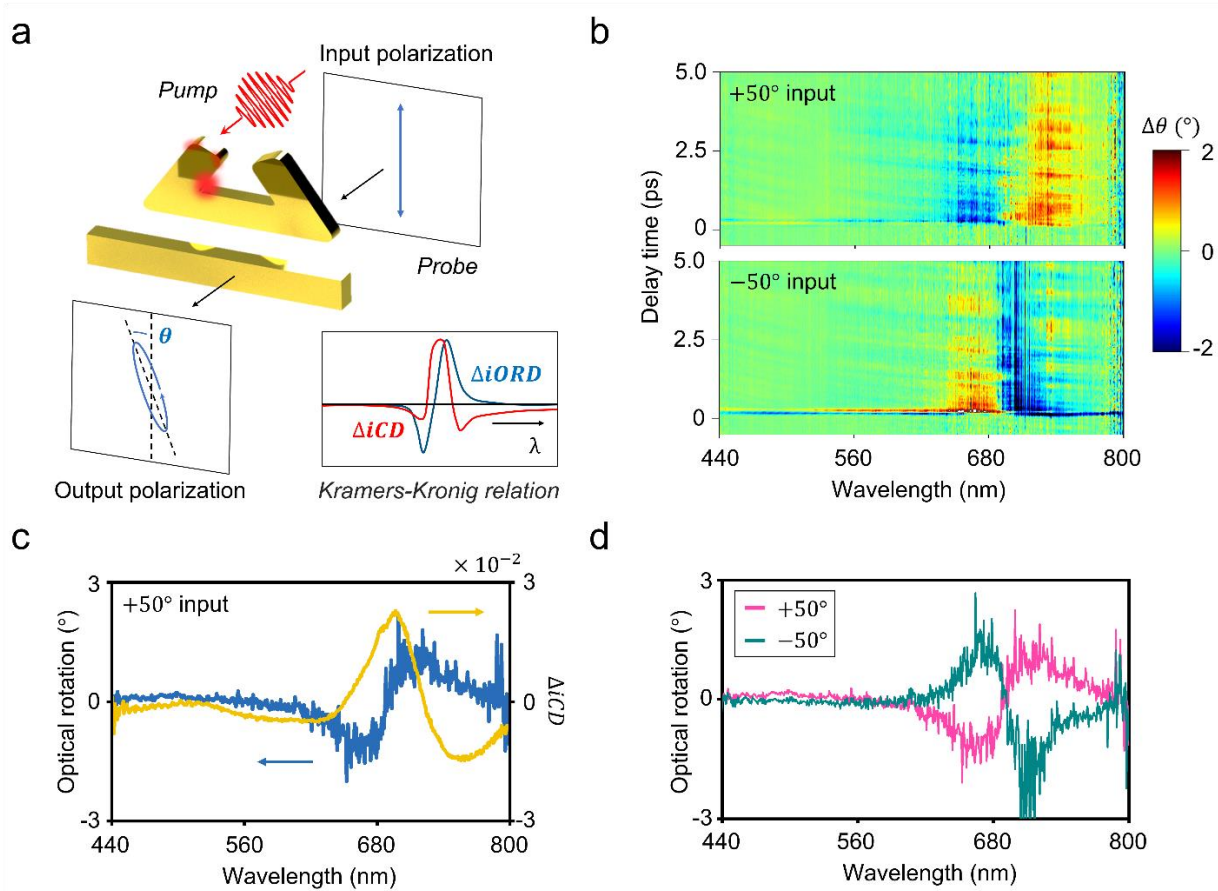

**Figure 4. Ultrafast optical rotation.** a) Conceptual schematic of hot-electron-induced optical rotatory dispersion (ΔiORD). b) 2D maps of measured induced optical rotation (Δθ) under +50° and -50° pump excitation acquired via Stokes polarimetry. c) Comparison of the spectral lineshape between maximum ΔiCD and maximum ΔiORD showing qualitative agreement as a Kramers-Kronig pair. d) Sign-flipped responses of ΔiORD for pump excitations with mirror-inverted polarization directions.

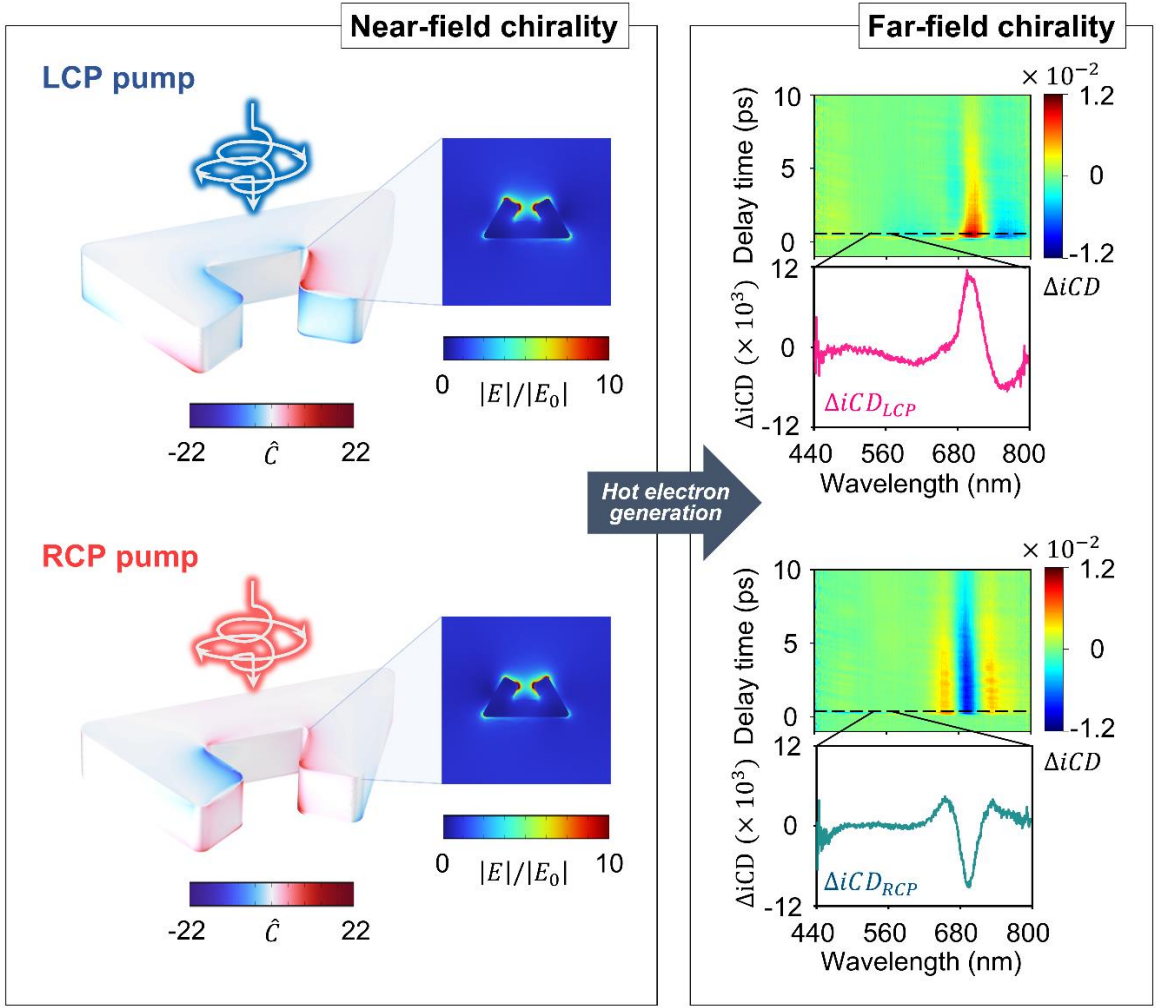

**Figure 5. Bridging near-field and far-field chiroptics.** a) Numerically simulated enhancement profiles of optical chirality and electric field of the achiral bilayer metasurface under LCP and RCP illumination showing chiral responses in the near-field. b) The near-field chirality produces spatially asymmetric hot-carrier populations, which leads to a far-field chiroptical response. The transient chiral responses induced by LCP and RCP pump display opposite signs which allow the chiral near-field distribution to be differentiated through far-field optics.